\documentclass{iau}
\usepackage{graphicx}

\title[ Questions on accreting mass and minimum magnetic field] 
{ Questions on  accreting mass and minimum magnetic field of millisecond pulsars} 

\author[C.M. Zhang]  
{C. M. Zhang
  }

\affiliation{National Astronomical Observatories, Chinese Academy
of Sciences, Beijing 100012, China \\ Email: {\tt
zhangcm@bao.ac.cn}}

\pubyear{2012}
\volume{291}  
\jname{\mbox{Neutron Stars and Pulsars: Challenges and Opportunities after 80 years}}
\editors{J.~van Leeuwen, ed.}
\begin{document}

\maketitle

\begin{abstract}
About 0.2 solar mass is  absorbed by the millisecond pulsar (MSP)
at the   binary accretion phase, while the polar magnetic field of
MSP is diluted to a magnitude of order $10^{8.5}$\,Gauss, which
is proportionally related to the mass accretion rate. It is found
that the minimum magnetic field of MSP can be as low as
$10^{7}$\,Gauss if the accretion rate of the binary system reaches its
the 
minimum value of $10^{15}$\,g/s. This bottom field has nothing to do
with the MSP initial field. Some questions on MPSs are proposed
and answered.

\end{abstract}


\firstsection 

\section{General picture of MSP evolution}
The millisecond  pulsar (MSP) is formed in the  binary accretion
phase,  while  the neutron star (NS) accretes matter from its
companion.  The magnetic field and spin period are both decreased
to the values of $10^{8.5}$\,Gauss and several milliseconds
(Wijnands \& van der Klis 1998), if the system absorbed the mass
of about 0.2 solar mass (\cite[Bhattacharya \& van den Heuvel
1991]{bha91}; \cite[Wang et al.\ 2011]{wang11};
\cite[Zhang et al.\ 2011]{zhang11}; \cite[Alpar et al.\ 1982]{alp82}; \cite[Tauris
2012]{tau12}).  Then the conventional NS is formed  from the
supernova explosion, while  a high magnetic field of about
$10^{11-13}$\,G  and  spin period of about 20-30 milliseconds,
which can be seen from the observed pulsar data of ATNF
(\cite[Manchester et al.\ 2005]{man05}). In  a NS
binary system,  if  the accretion mass of $\sim 0.01 M_{\odot}$ is
accreted from the companion, a NS can  be spun up to several tens
of  milliseconds, while its magnetic field will be decayed  to
$\sim 10^{9-10}$\,G, e.g. the double pulsar system (Lyne et al.
2004; van den Heuvel 2004).  Therefore, more mass accreted and
less magnetic field of NS achieved (\cite[Zhang \& Kojima
2006]{zhang06}).  In this short paper we present some enquiries and basic
conclusions on MSP evolution.

\section{ The bottom field and minimum field of MSP}

  During the accretion phase,  if the NS field is high,
  then the accreting MHD materials  flow  along  the polar field
  lines to fall on the NS surface, while the MHD matter is
  piled-up at the magnetic polar cap and it flows from the polar
  cap to the equator, while the MHD flow drags the field lines
asides to dilute the polar field strength (\cite[Zhang \& Kojima
2006]{zhang06}). The accretion flow continues until the field
lines are all trapped into the star, at where the magnetosphere of
NS equals to the NS radius, which can give a bottom field
($B_{f}$) of NS of about $\sim 10^{8.5}$\,G after the system
accretes about 0.2 solar mass,
\begin{equation}
B_{f}   \simeq  10^{8.5} (G) (\dot{M}/10^{18 } g/s)^{1/2},
\end{equation}
where $\dot{M}$ is the accretion rate.  If the accretion rate is
at its minimum value of $10^{15}$ g/s, then the minimum field
strength of MSP is obtained as
\begin{equation}
B_{min}   \simeq  10^{7} (G) (\dot{M}/10^{15} g/s)^{1/2}.
\end{equation}
From the known derived fields of MSPs, we find the minimum value
of field is about $10^{7.5}$\,G (\cite[Manchester et al.\ 2005]{man05}),   which is very close to our theoretical
result. Moreover, the field and spin evolutions with the accreted
mass can be given, and both   decay until the   bottom values
after   system accretes about 0.2 solar mass.

\section{Does NS field decay after the accretion phase?}

Generally, from a long-term point of views, the field decays while
the accreted mass is added, and field has little decay if the
accretion phase stops. The evidence for this can be found from the
binary pulsar system (NS+white dwarf), where the temperature of WD
is observed that implies a cooling age of system, which gives a
conclusion that the field has little decay at the time scale of
$10^{9}$ yr if there is no accretion.

\section{The accretion must result in the field decay of NS?}
From the accretion induced field decay model (\cite[Zhang \&
Kojima 2006]{zhang06}), if the accretion MHD dragging at the polar
cap is totally frozen, then all accreted matter contributes to the
field decay. In case the MHD frozen efficiency is not 100\%, for
instance 1\%, then 0.2 solar mass accretion only makes field decay
to $10^{10}$\,G. If this efficiency is as low as 0.01\%, then 0.2
solar mass accretion makes the little field decay.  The frozen
efficiency may be related to the magnetic inclination angle of NS,
then detail of which is still in consideration. Say, some NSs may
have little field decays after accreting 0.2 solar mass.

\section{The role of accretion rate on MSP field}

The accretion mass has a dominant role in MSP evolution; the
accretion rate is also a factor. From X-ray NS, the average rate
is about $\dot{M}/10^{17}$ g/s.  If the rate is less than
$\dot{M}/10^{15}$ g/s,  then the Ohmic decay velocity will be
faster than  the accretion deepen-in velocity, while the accretion
cannot bury the field into the NS core region. Therefore, a
minimum accretion rate is needed for the field decay. The high
accretion rate corresponds to a high bottom field, e.g.
Eddington rate will correspond to a bottom  field of $10^{8.5}$\,G.

\section{The mass of MSP: 1.6 solar mass on average }
Two decades ago, researchers often thought that the mass of MSP should
be usually as high as 2.0 $M_{\odot}$. However, the recent statistics
of 65 measured NS masses indicates that the average mass of MSP is
only 1.6 solar mass (\cite[Zhang et al. 2011]{zhang11}), comparing to
the mass of slow rotating NS of about 1.4 solar mass, and MSP seems to
absorbs only 0.2 solar mass in average. Then a question arises where
it goes the one solar mass of WD companion?  Does an MSP only accrete
20\% mass of its companion? After the spinning-up of an MSP, some
accreted mass is finally rejected from the NS?  The EOS of MSPs is
different from that of normal NSs? These questions are still open.

\section{Alternative possibility: MSP formation from AIC? }
MSPs are usually thought to be formed by a
spin-up of NS in a binary system by accretion. Then, the idea
that an MSP forms in an accretion induced collapse of a WD is not
automatically excluded.   What is a specialty of an MSP from AIC?
The mass of an MSP by AIC should be less than Chandrasekhar 
limit, e.g. 1.3 solar mass. From the measured masses of MSPs
(\cite[Zhang et al. 2011]{zhang11}), it really exist
3 MSPs with the masses less than 1.3 solar masses.  Then we cannot be
sure if these low mass MSPs are formed in the low mass states,
e.g. one solar mass, or by AIC processes. More MSP
masses are needed, by which we can distinguish the details of MSP
mass and its formation processes.

\section{Bottom magnetic field of MSP has no relation to its initial field}
The evolution history of MSPs can be understood in this way: the
initial magnetic field of NS can be as high as $B \sim
10^{13}$\,G, and it can decay to $B \sim 10^{8}$\,G, after
accreting $\sim 0.2 M_\odot$, while NS magnetosphere is the same
size of NS. In other words, the bottom field of MSP is determined
by the condition that the radius of  NS magnetosphere equals the
radius of NS. This fact means that the bottom field is nothing to
do with its initial  field strength!   From the field and spin
period  distributions of MSPs and pulsars (\cite[Wang et
  al. 2011]{wang11}),  we notice that the field distribution 
of normal pulsars ranges at $B \sim 10^{11-13}$\,G, then that of
MSPs ranges narrowly at $B \sim 10^{8-9}$\,G, which should be the
effect of bottom field, at where the fields stops but is no
relation to its born field.  The significant application of bottom
field is that the high and low luminosity X-ray sources, Z and
Atoll, share the similar kHz QPO frequencies, indicating the
similar magnetosphere radii at scale of NS radii (Zhang 2004;
Zhang et al. 2006)

\section{Magnetic structure of MSP: strong magnetic domain $10^{14-15}$\,G?}

If accretion drives decays of the magnetic field of an MSP, is
the magnetic structure altered too? Yes;
the polar
field lines of MSP are dragged to the equator region, where all field
lines are sinked into the core region of NS \cite[(Zhang \& Kojima
  2006)]{zhang06}. Thus,  
the estimation indicates that the polar field of MSP is as low as
$10^{8}$\,G, the equator field, beneath the NS surface,  can be as
high as $10^{13-14}$\,G, and the core field of NS may be even
higher that $10^{14}$\,G, e.g. $10^{15}$\,G. In other words, the MSP
magnetic structure is redistributed after accretion. The leaked-out
field lines of the MSP may produce the effect that the local field 
at the magnetic equator should be much higher than $10^{8}$\,G,
e.g. $10^{11-12}$\,G, then the global field of MSP is dominated by
a polar field with the low value $10^{8}$\,G. Radio observations
of MSPs should reflect their low field characteristic, while the
X-ray observation may show their local strong fields. The
complex magnetic structure of MSP may be noticed from the high
energy emissions, e.g. {\it Fermi}, in the future.

~\\
{\bf Acknowledgements:}
 This work has been supported by National
Basic Research Program of China (2012CB821800), NSFC10773017 and
NSFC11173034.


\begin{thebibliography}{}


\bibitem[Alpar et al. (1991)]{alp82}
{Alpar M. A., Cheng A. F., Ruderman M. A., Shaham J.}, 1982, \textit{Nature}, 300, 728

\bibitem[Bhattacharya \& Heuvel (1991)]{bha91}
{Bhattacharya D. \& van den Heuvel E. P. J.} 1991, \textit{Phys. Rep.}, 203, 1



\bibitem[Lyne, Burgay \& Krammer et al. (2004)]{lyn04}
{Lyne A. G., Burgay M. \& Kramer, M. et al.}, 2004, \textit{Science}, 303, 1153.




\bibitem[Manchester, Hobbs G., Teoh \& Hobbs M. (2004)]{man05}
{Manchester R. N., Hobbs G. B., Teoh A., \& Hobbs M. }, 2005,  \textit{AJ}, 129, 1993



\bibitem[Tauris (2012)]{tau12}
{Tauris M.}, 2012, \textit{Science}, 335, 561



\bibitem[van den Heuvel (2004)]{heu04}
{van den Heuvel E. P. J.}, 2004, \textit{Science}, 303, 20


\bibitem[Wang et al. (2011)]{wang11}
{Wang, J., Zhang, C. M., Zhao, Y. H. et al.}, 2011, \textit{A\&A}, 526, 88


\bibitem[Wijnands \& van der Klis (1998)]{wij98}
{Wijnands R. \& van der Klis M.}, 1998, \textit{Nature}, 394, 344


\bibitem[Zhang (2004)]{Zhang04}
{Zhang, C.M.} 2004, \textit{A\&A}, 423, 401
%


\bibitem[Zhang \& Kojima (2006)]{zhang06}
{Zhang C. M., Kojima Y.}, 2006, \textit{MNRAS}, 336, 137

\bibitem[Zhang et al. (2006)]{Zhang06}
{Zhang, C. M., Yin, H. X., Zhao, Y. H. et al.} 2006,
\textit{MNRAS}, 366, 1373

\bibitem[Zhang, Wang \&Zhao (2011)]{zhang11}
{Zhang C. M., Wang J., Zhao Y.H. et al.}, 2011, \textit{A\&A}, 527, 83


\end{thebibliography}
\end{document}